\documentclass[9pt,twocolumn,twoside,superscriptaddress]{revtex4-2}

\usepackage{siunitx}
\usepackage{graphicx}
\usepackage{amsmath,amsfonts}
\usepackage{xcolor}

\newcommand{\QFI}{\mathcal{F}}
\newcommand{\QFIn}{\tilde{\QFI}}
\begin{document}

\title{Continuous wave multi-pass imaging flow cytometry}
\author{Yonatan Israel}
\affiliation{Physics Department, Stanford University, Stanford, California 94305, USA}
\author{Joshua L. Reynolds}
\affiliation{Applied Physics Department, Stanford University, Stanford, California 94305, USA}
\author{Brannon B. Klopfer}
\affiliation{Applied Physics Department, Stanford University, Stanford, California 94305, USA}
%\affil[*]{Corresponding email: bklopfer@stanford.edu}
\author{Mark A. Kasevich}
\affiliation{Physics Department, Stanford University, Stanford, California 94305, USA}
\affiliation{Applied Physics Department, Stanford University, Stanford, California 94305, USA}

\begin{abstract}
We present a wide-field multi-pass implementation of label-free imaging flow cytometry. Our technique is shown for high-speed flow imaging of ensembles of human red blood cells with up to four passes, demonstrating x4 enhancement in contrast and signal-to-noise. We show that our technique approaches close to the quantum limit of measurement sensitivity, extending the range of optimal imaging to samples in the weakly absorbing regime. This allows for near optimal imaging sensitivity and throughput in a practical scenario of imaging a dynamic sample under limited illumination intensity, surpassing the sensitivity achieved with currently available quantum light sources.

%continuous-wave implementation of a wide-field optical multi-pass microscope.
%It can be operated with a spatially and temporally incoherent light source, and requires no active outcoupling or exotic detection schemes.
%This implementation is capable of deterministically interrogating a sample sequentially up to $m=4$ times.
%Through multiple interrogations, a linear enhancement in phase shift and absorption imparted by the sample can be achieved, fundamentally increasing the signal-to-noise of the obtainable images.
\end{abstract}

\maketitle
%%%%%%%%%%%%%%%%%%%%%%%%%%%%%%%%%%%%%%%%%%%%%%%%%%%%%%%%%%%%%%%%%%%%%%%%%%%%%%%%
% Begin body
%%%%%%%%%%%%%%%%%%%%%%%%%%%%%%%%%%%%%%%%%%%%%%%%%%%%%%%%%%%%%%%%%%%%%%%%%%%%%%%%

\section{Introduction}
Label-free imaging techniques measuring weak optical properties, e.g., the phase shift and absorption due to a thin biological specimen, are typically limited in sensitivity. In the common case where instrumental and environmental noises are negligible, such optical imaging techniques are fundamentally limited by photon shot-noise \cite{PhaseMicLimits2016,PhasemicroscopyReview2018}. 
%places some fundamental and practical limits on the imaging method performance.
By utilizing quantum properties of light, quantum imaging protocols have been proposed and realized for various label-free imaging techniques in order to enhance their sensitivity \cite{QsensReview2018,QimagReview2019}, for example in phase microscopy \cite{NOONmicroscopy2014}, absorption microscopy \cite{QlAbs2017}, and nonlinear microscopy \cite{Warwick_NL_Nature2021,QSBS2022}. While these techniques promise enhanced sensitivity, their implementation has so far remained limited and impractical compared to classical imaging techniques.

Multi-pass microscopy is an alternative route to enhancing sensitivity that has recently been investigated  \cite{mpm,oam}. In multi-pass imaging, the optical probe field transits the image target multiple times, increasing the phase shift and absorption of the probe field, which results in an increase in the signal-to-noise ratio (SNR). In this work, we demonstrate that a multi-pass approach can enhance the image sensitivity in a practical scenario of shot-noise limited imaging of a dynamic flow sample. 

Such dynamic samples are common in biological imaging, where the objects being imaged either evolve over time or move quickly through the field-of-view of the imaging system. In particular, imaging flow cytometry uses fast flow rates for high throughput characterization and sorting of large cell populations in biological research and clinical diagnostics  \cite{fast_flow00,Goda2012, Doan2018}. In such scenarios, the imaging speed must be faster than the sample dynamics to avoid blur, which, for a limited illumination source intensity, results in limited imaging sensitivity due to shot-noise. %Here we show that multi-pass imaging allows increasing the sensitivity of imaging flow cytometry. 

In this work, we describe a wide-field multi-pass imaging approach that enhances the sensitivity of label-free imaging flow cytometry. While previous multi-pass microscopes relied on pulsed illumination and time-gated cameras to form the multi-pass image \cite{mpm,oam}, our implementation is compatible with continuous-wave (CW) illumination and conventional cameras.
Furthermore, it can accommodate spatially and temporally incoherent sources, suppressing spurious contributions from out-of-focal-plane scattering, and can achieve up to four passes through the sample. While scanned-cavity methods, such as those described in Refs. \cite{motschCavityenhancedRayleighScattering2010,changCavityQEDAtomic2012}, can achieve significantly greater signal enhancements, they cannot operate with incoherent illumination or in  wide-field, which allows substantially faster image acquisition times. %\jrcomment{Feel like we need a sentence in this paragraph that more concretely states that the demonstrated benefits are good for imaging flow cytometry.}

\begin{figure*}[!htbp]
	\centering
	\includegraphics[width=\textwidth]{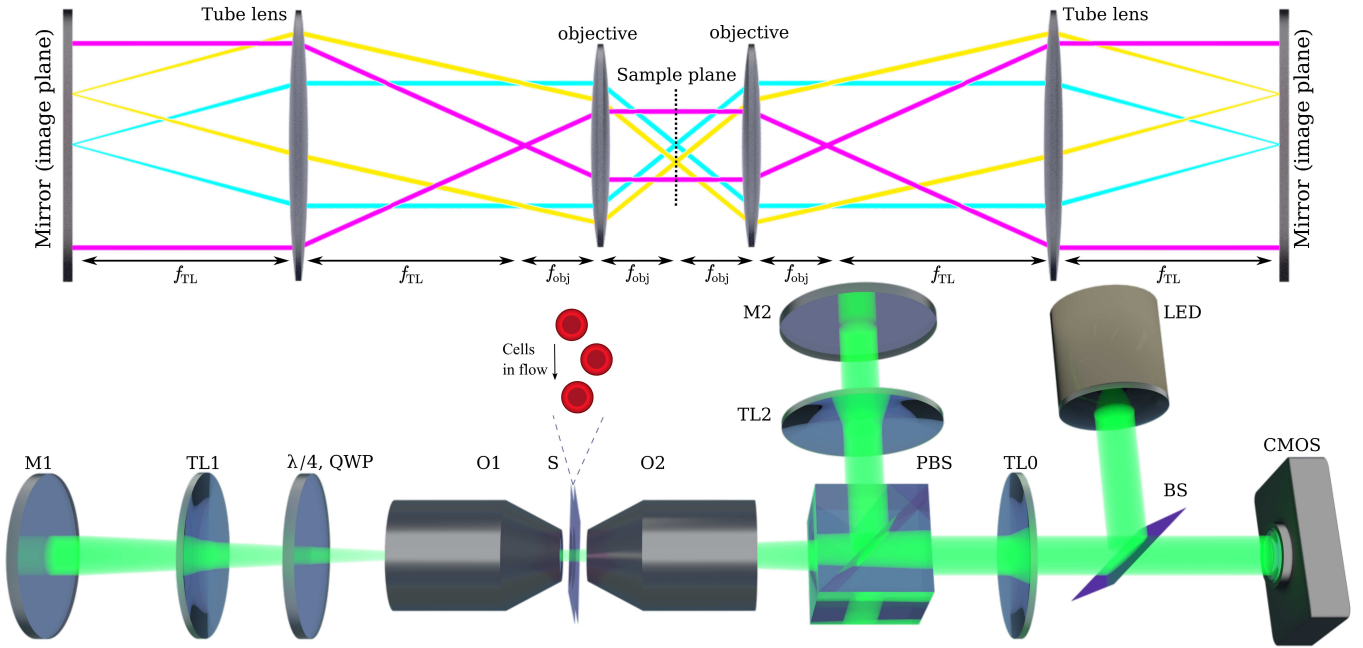}
	\caption[setup]{
	Multi-pass imaging schematics. {\bf Top:}
	ray trace of a multi-pass microscope with tube lens and objective focal lengths $f_\text{TL}$, $f_\text{obj}$, respectively.
	Rays are shown for plane wave illumination (magenta), on-axis point scatterer (cyan), and off-axis point scatterer (yellow).
	{\bf Bottom:}
	optical setup, consisting of two microscopes placed between end mirrors M1 (2). 
	The microscopes consist of NA$=0.8$ objectives (O1 and O2) and a tube lens (TL1 and TL2).
	The light (LED) is in- and outcoupled via a beamsplitter (BS) and a polarizing beamsplitter (PBS).
	A quarter-waveplate ($\lambda/4$, QWP) is used to switch between $m=2$ and $4$ interrogations.
	$m = 1$ is obtained by severely defocusing the left-hand objective O1.
	For clarity, additional extracavity optics are omitted.
	%\jrcomment{Is the dashed triangle going into the sample plane tilted?}
	}
	\label{fig:setup}
\end{figure*}

% These just make life easier:
\newcommand{\Ei}{\ensuremath{E_\text{i}}}
\newcommand{\Es}{\ensuremath{E_\text{s}}}
\newcommand{\Is}{\ensuremath{I_\text{s}}}
\newcommand{\Er}{\ensuremath{E_\text{r}}}
\newcommand{\Ir}{\ensuremath{I_\text{r}}}
\newcommand{\Id}{\ensuremath{I_\text{d}}}
\newcommand{\tavg}[1]{\overline{#1}}

\section{Methods}

The basic CW multi-pass configuration is illustrated in Fig. \ref{fig:setup}.  In this implementation, the imaging beam from an incoherent LED transits a self-imaging optics path either two or four times, depending on the optics configuration. 

For a two-pass implementation, illumination light is incoupled through a non-polarizing beamsplitter (BS).  The illumination field is focused in the sample region with objective O2. %\yicomment{OBJ 1 <-> 2?} \jrcomment{sounds good to me}
After transiting the sample, the light is collected by objective O1.
These fields are then imaged by a tube lens TL1 onto a reflective planar mirror M1.
By symmetry, this reflected image is reimaged onto the sample when appropriately aligned.
Ray-traces of the relevant optical paths are also shown in Fig. \ref{fig:setup}. In order to tune the focus, both the O1 objective and sample stage are actuated along the optical axis with encoded piezo stages.

A four-pass implementation utilizes a polarizing beam splitting cube (PBS) and a quarter-wave plate ($\lambda/4$, QWP).
In this case, the linear polarization of the beam is rotated by $\pi/2$ as it passes twice through QWP, resulting in redirection of the retroreflected field (after having passed twice through the sample) onto a planar mirror M2.  This mirror reflects the fields back into the microscope optical path as in the two-pass implementation. 
After two more passes through the QWP (four total passes through the sample), the beam exits the microscope and enters the image-forming optics.
We toggle between the two-pass and four-pass configurations by adjusting QWP: rotating the polarization of the retroreflection leads to 4 passes, while not rotating it leads to 2 passes.
We note that for high NA CW illumination of an isotropic sample, the maximum number of passes is limited to four since the polarization has rotated by $\pi/$2 for every round trip through the re-imaging cavity.

To operate the apparatus in an $m=1$ configuration (i.e. without multi-passing the sample), we operate in the $m=2$ configuration and severely defocus the objective O1 (see Fig. \ref{fig:setup}).
In this configuration the sample is re-illuminated with a defocused image of the sample, and only objective O2 forms an image on the camera.
Thus, we form all images at $m=1,2$ and $4$ using an identical light source (Thorlabs M530F2), CMOS camera (Teledyne FLIR BFS-U3-32S4M-C), objectives (Zeiss, Plan-Apochromat 20x/0.8), and  other optics.

\section{Sensitivity analysis}
\subsection{Signal-to-noise}
To describe the SNR in bright-field multi-pass imaging, assuming in the following an absorption-only sample (ignoring other optical losses), we use the average number of detected photons:
\begin{equation}
    \left<\hat{n}(x,y)\right> = \left<\hat{n}(x,y)\right>_\text{in} \cdot\eta^m(x,y),\label{eq:I}
\end{equation}
where $\left<\hat{n}(x,y)\right>_\text{in}$ is the average number of photons at the apparatus input, $\eta(x,y)$  and $1-\eta(x,y$) are the optical transmission and absorption of the specimen, respectively, at transverse positions $x$ and $y$, and $m$ is the number of passes.
For a measurement at the shot-noise limit, the image SNR is then given by 
\begin{equation}
    \text{SNR}_m = \frac{|\left<\hat{n}\right>_\text{in}-\left<\hat{n}\right>|}{\sqrt{\left<\hat{n}\right>_\text{in}+\left<\hat{n}\right>}} = \frac{\sqrt{\left<\hat{n}(x,y)\right>_\text{in}} \cdot (1-\eta^m)}{\sqrt{1+\eta^m}}.\label{eq:snr_abs}
\end{equation}
For weakly absorbing (high transmissivity) samples, the image SNR is
\begin{equation}
    \label{eq:snr_fin}
    \text{SNR}_m \approx  \sqrt{\frac{\left<\hat{n}\right>_\text{in}}{2}} \cdot \alpha z \cdot m,
\end{equation}
where $\alpha$ is the sample absorption per unit depth, related to the transmission by $\eta$ = $e^{-\alpha z}$, and $z$ is the sample thickness.
The SNRs of absorption measurements in multi-pass imaging are thus enhanced by a factor $m$. We note that for imaging at fixed SNR, the required number of detected photons decreases by a factor of $m^2$. While we describe here the simple case of an absorption-only sample, multi-pass imaging can also be used to enhance phase contrast as well (see Supplement 1).

\subsection{Fisher Information}
\begin{figure}[h!]
	\centering
	\includegraphics[width=\columnwidth]{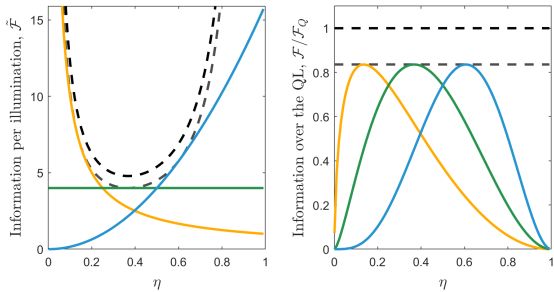}
	\caption[FI]{
    Limits to absorption information. {\bf Left:} information per illumination, and {\bf Right:} information relative to the quantum limit (QL) , for single-pass (gold), two passes (green), and four passes (blue). The quantum $\QFIn_Q$ (dashed black) and the classical limit $\QFIn_{m_{opt}}$ (dashed gray) on information are show for comparison. 
	\label{fig:FI}}
\end{figure}

Next, we calculate the amount of information about the sample that is carried out by the probe light. We evaluate the effectiveness of imaging flow cytometry for multi-pass setups and compare it with the quantum limit for information per illumination intensity. The information contained on $\eta$ in %{the intensity measurement $I$ and in} 
the quantum state of light $\rho_{\eta}$ is denoted as %{$\FI$\, for the Fisher information (FI) and} 
$\QFI$\, and known as the quantum Fisher information (QFI) %, respectively
\cite{Helstrom,Holevo1982,Braunstein1994}. The QFI bounds the measured variance of the absorption according to %the Crame\'r-Rao bound (CRB) and 
the quantum Crame\'r-Rao bound (QCRB),
\begin{equation}
     {\QFI(\rho_\eta)}\underset{\text{QCRB}}{\geq} 
     %{F(\rho_\eta,T)}\underset{\text{CRB}}{\leq}
     1/\text{Var}(\eta).
    \label{eq:CRB}
\end{equation}
%where $T = [t_i]$ is the positive operator-valued measure (POVM) associated with the measured probabilities $p(i|\eta) = tr(t_i\rho_{\eta})$. In \autoref{eq:CRB}, the CRB relates the variance of the unbiased estimate $\text{Var}(\eta)$ to the FI, defined as $F(\rho_{\eta},T) = \sum_i p(i|\eta) (\partial_{\eta} \ln p(i|\eta))^2 $, while the QCRB relates the FI to the its maximum value $\QFI$, the QFI, obtained by maximizing over all POVMs \cite{Braunstein1994}. %\jrcomment{I see your commented out comment below. I like this introductary paragraph to FI. My only substantial comment on the whole section is that while an analytic form for \textit{F} is given in the previous sentence, the QFI is given only a more qualitative description. Might be more parallel and/or helpful to give either quantitative or qualitative descriptions for both. I lean towards just providing qualitative descriptions for both and maybe putting quantitative in the supplement, and, if we're going to calculate QFI in the supplement, maybe we should supply a more general equation for the QFI there.} %tr(\rho_{\eta}\mathcal{L}_{\eta}^2(\rho_{\eta}))$, where $mathcal{L}_{\eta}(\rho_{\eta})$ =  \yicomment{maybe all the above in redundant}. 
To compare different experimental strategies we use the QFI per (input) illumination average photon number $\tilde{\QFI} = \QFI/\langle\hat{n}\rangle_{in}$. Using classical light, for multi-pass we have \cite{JM_abs_PRR2020,JM_PhaseAbs_PRL2020}
\begin{align}
& \QFIn_{m} = m^2\eta^{m-2}. \label{eq:QFI-MP}
\end{align}
In particular, the single-pass QFI limited by the (classical) illumination average photon number is given by $\QFIn_{1} = 1/\eta$. We note that in Eq. \ref{eq:QFI-MP}, the QFI bounding $\text{Var}(\eta)$ scales with the number of passes similarly to $1/\text{SNR}^2\sim 1/\left<\hat{n}\right>_\text{in}$ of Eq. \ref{eq:snr_fin}. Furthermore, we show in Supplement 1 that the variance in $\eta$ using intensity measurements in our multi-pass setup (Eq. \ref{eq:I}) saturates the QFI for multi-pass using classical light $\QFIn_m$ (Eq. \ref{eq:QFI-MP}). By permitting $m$ to be continuous, we find the maximal multi-pass QFI $\QFIn_{m_{\text{opt}}} = 4/(e\cdot\eta\ln{\eta})^2$ for $m_{\text{opt}} = -2/\ln{\eta}$. %\jrcomment{Maybe would flip, we find that the QFI is maximized as ... for an optimal number of passes $m_{opt}$.}
Note that non-sample losses are ignored, which would reduce the QFI by a factor depending on these losses \cite{JM_abs_PRR2020,JM_PhaseAbs_PRL2020}.

To compare the capability of the above techniques, which use classical light, with the absolute upper limit allowed by using quantum light as well as multi-pass approaches, we derive the quantum limit (see Supplement 1),
\begin{align}
    \QFIn_Q \approx %-\frac{\mathcal{W}(2+\mathcal{W})}{(\eta\ln{\eta})^2}=
    \frac{0.65}{(\eta\ln{\eta})^2}.    
    \label{eq:QFI_Q}
\end{align}
It is interesting to note that under constant illumination, an optimal quantum probe allows for a maximal reduction in variance over an optimal classical multi-pass strategy by $\text{Var}(\eta)_Q/\text{Var}(\eta)_{Cl} = \QFIn_{m_{\text{opt}}}/\QFIn_Q=0.84$. Quantum light sources are typically orders of magnitude less intense than classical light sources, and quantum approaches are, therefore, currently less sensitive than classical ones. For low-intensity illumination of damage-sensitive samples, quantum light has enabled at most a $\times 1.4$ performance enhancement over single-pass classical strategies \cite{NOONoverSNL_NPh2017,Warwick_NL_Nature2021}. Furthermore, it was recently shown that classical light in a ring resonator can achieve the quantum limit for absorption measurement \cite{JM_cavity_PRL2022}. While such an approach is not directly applicable for wide-field imaging using incoherent light, an optimized cascaded interferometer could potentially extend the sensitivity of multi-pass and achieve the quantum limit in imaging using classical light \cite{JM_LossyPhase_PRA2017,koppell2022Arxivl}. 

Fig. \ref{fig:FI} shows the QFI under fixed illumination intensity and the QFI over the quantum limit for single-pass and multi-pass. While single-pass measurements approach the classical limit, or 84\% of the quantum limit for strongly absorbing signals with $\eta\approx 0.13$, four-pass measurements approach these limits for weakly absorbing samples with $\eta\approx0.61$ ($\eta^m\approx0.13$). Furthermore, four-pass still maintain a 16-fold increase in information over single-pass for the weak absorption limit $\eta\rightarrow1$. 

\begin{figure}[b!]
	\centering
	{\includegraphics[width=\columnwidth]{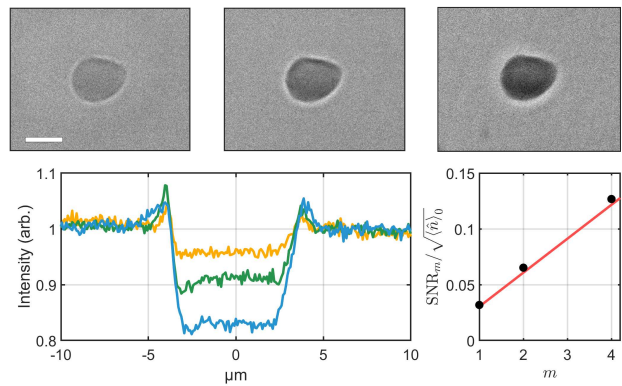}
	\caption[HRBC]{
	SNR enhancement for a single static human red blood cell (HRBC). 
	{\bf Top:} 
	    bright-field micrographs of a single HRBC at (left to right) $m=1$, 2, and 4. The images are normalized to the average background squared-root photon flux for each $m$. Scale bar is \SI{5}{\micro\metre}.
	{\bf Bottom left:}
		lateral cross sections at $m=1$ (gold), 2 (green), and 4 (blue) showing contrast enhancement.
	{\bf Bottom right:}
	    SNR after $m$ passes, normalized to the average squared-root photon flux without the object, compared with Eq. \ref{eq:snr_fin} for $\eta$ = 0.957.
	}
	\label{fig:hrbc}}
\end{figure}

\section{Results \& Discussion}
We first demonstrate multi-pass enhancement with a static sample.
Micrographs of a human red blood cell (HRBC), normalized to the average background photon count at $m=1$, 2, and 4 are shown in Fig. \ref{fig:hrbc}, along with lateral cross sections, which are averaged over 20 camera pixels ($\approx \SI{1.7}{\micro\meter}$ in the sample plane) along the vertical image direction. %(\SI{1}{px} $\approx$ \SI{84}{\nano\meter}). 
The gray intensity scale is constant across the three images, and the contrast enhancement is readily visible. 
%The contrast enhancement is readily visible (the color scales are held constant for all images).
The cross sections display approximately linear buildup in contrast as a function of the number of interrogations, as expected for absorption contrast.
The measured SNRs use the average squared-root photon flux without the object $\left<\hat{n}\right>_0$, which accounts for optical losses other than in the sample, instead of the average input photon number $\left<\hat{n}\right>_\text{in}$ used in Eq. \ref{eq:snr_fin}.
The SNR normalized by $\sqrt{\left<\hat{n}\right>_\text{0}}$, SNR$_m/\sqrt{\left<\hat{n}\right>_\text{0}}$, is averaged over each HRBC image; as shown in Fig. \ref{fig:hrbc}, it increases linearly with $m$, consistent with Eq. \ref{eq:snr_fin}.
We estimate the average HRBC transmission shown in Fig. \ref{fig:hrbc} to be $\eta$ = 0.957, while the average SNR$_m$ is measured to be 0.99, 2.1, and 3.6 at $m=1$, 2, and 4, respectively. The sub-linear enhancement in SNR results from optical losses (see Supplement 1). 

\begin{figure*}
	\centering
	{\includegraphics[width=18cm]{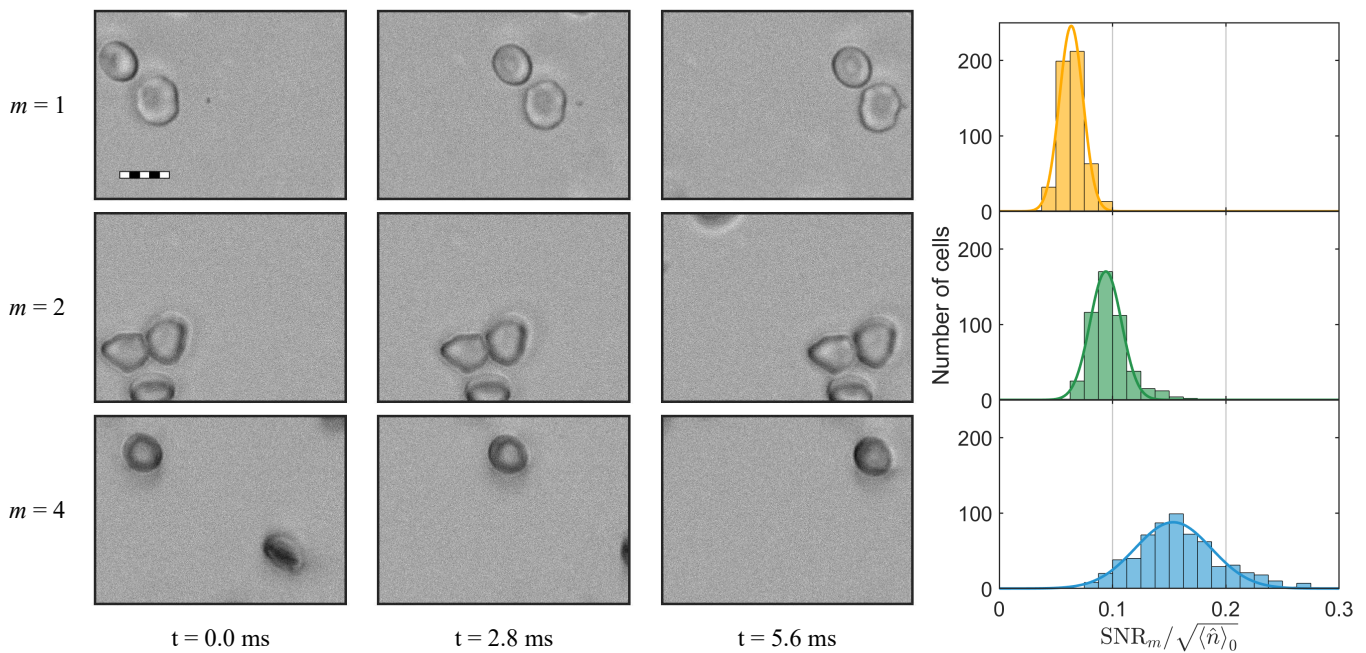}}
	\caption{Signal enhancement using multi-pass for red blood cells in flow. 
	{\bf Left:}
	    HRBCs flowing left to right at $m=1$, 2, and 4 (top to bottom). 
	    The spatial and color scales are constant for all images (\SI{10}{\micro\meter} scale bar with \SI{2}{\micro\meter} ticks) to show enhancement in the absorption contrast. 
	    Flow rates are approximately \SI{5}{\milli\meter\per\second}. 
	    The camera exposure time is \SI{10}{\micro\second}, 360Hz frame rate, and all pixels are normalized to their average value measured in the absence of cells.
    {\bf Right:}
        histograms with Gaussian fits showing the average SNR$_m/\sqrt{\left<\hat{n}\right>_\text{0}}$ for cells in flow. 
        The signal includes a constant contribution from phase contrast and a contribution from absorption contrast that is enhanced by $m$. 
        The sensitivity enhancement results in a broadening of the distributions. 
	}
	\label{fig:flow}
\end{figure*}

When the image acquisition time and illumination intensity are limited, as is the case of imaging flow cytometry, the image SNR is limited. For a $\SI{1}{mW}$ source illuminating a sample in wide-field and detected using a few mega-pixel camera within an $\SI{10}{\micro\second}$ exposure time, each pixel records $\left<\hat{n}\right>_\text{0} \approx 1000$ photons. For such values of $\left<\hat{n}\right>_\text{0}$, the SNR in single-pass imaging is $\text{SNR}_1\approx1$ (see Eq. \ref{eq:snr_abs}) for a weakly absorbing sample having $\eta\approx0.95$.
This SNR can be increased by a factor of 4 by multi-passing a probe with fixed illumination. Alternatively, the required acquisition time necessary to image at fixed SNR would decrease by a factor of 16 for $m=4$ over a single-pass measurement for a fixed illumination intensity.

To demonstrate the applicability of multi-pass to high-throughput applications such as imaging flow cytometry, we imaged HRBCs flowing through a glass rectangular capillary with a channel depth of \SI{20}{\micro\meter}. 
Sample cells in flow are shown in Fig. \ref{fig:flow}, where the photon count in each pixel $\left<\hat{n}\right>$ is normalized to its average value measured in the absence of flowing cells $\left<\hat{n}\right>_\text{0}$, and the gray intensity scale is the same across all images. 
For our depth-of-field, both in-focus and out-of-focus cells are visible across the channel; however, only in-focus cells are included in our analysis.

The images shown in Fig. \ref{fig:flow} were taken under \SI{1}{\milli\watt} illumination, limited by our fiber-coupled LED source intensity, with exposure times of \SI{10}{\micro\second}, and show flow rates of approximately \SI{5}{\milli\meter\per\second}. 
We note that the above mentioned flow rate was kept low for demonstration of the imaging flow over a few frames. 
With this setup, we have imaged cells flowing at speeds of up to approximately \SI{60}{\milli\meter\per\second}, which is comparable with flow rates of commercial high-speed imaging flow cytometers \cite{ImageStream}, using the full field of our camera with similar SNRs for the same acquisition time. 
Under these conditions, the maximum motion-induced blur is \SI{600}{\nano\meter}, comparable to the diffraction limit in our imaging system. 

%Histograms of the measured average SNR over the squared root of the illumination photon number SNR$_m/\sqrt{\left<\hat{n}\right>_\text{0}(x,y)}$ \yicomment{SNR$_m/\sqrt{\left<\hat{n}\right>_\text{0}}$} at $m=1$, 2, and 4 are shown in \autoref{fig:flow}, along with sample cells in flow. \yicomment{Move SNR to after the images?}
%In the images shown in \autoref{fig:flow}, the photon count in each pixel $\left<\hat{n}\right>$ is normalized to its average value measured in the absence of flowing cells \yicomment{$\left<\hat{n}\right>_\text{0}$}, and the images are shown with the same gray intensity scale.
%For our depth-of-field, both in-focus and out-of-focus cells are visible across the channel; however, only in-focus cells are included in our analysis.
%The images shown in \autoref{fig:flow} were taken under \SI{1}{\milli\watt} illumination, limited by our fiber-coupled LED source intensity, with exposure times of \SI{10}{\micro\second}, and show flow rates of approximately \SI{5}{\milli\meter\per\second}. 
%We note that the above mentioned flow rate was kept low for demonstration of the imaging flow over a few frames. 
%With this setup, we have imaged cells flowing at speeds of up to approximately \SI{60}{\milli\meter\per\second}, which is comparable with flow rates of commercial high-speed imaging flow cytometers \cite{ImageStream}, using the full field of our camera with similar SNRs for the same acquisition time. 
%Under these conditions, the maximum motion-induced blur is \SI{600}{\nano\meter}, comparable to the diffraction limit in our imaging system. 

Here, an enhancement of image absorption contrast is readily visible as well, with thicker cell areas having higher absorption than the characteristic concave centers of the HRBCs. Phase contrast is apparent at cells' edges and centers. However, the phase contrast enhancement with $m$ is limited by the axial depth of the channel. For example, overfocused planes of the sample imaged by O1 are underfocused by O2, washing out the phase contrast that would be obtained from imaging out-of-focus planes of a sample using a single objective. The phase contrast is therefore constant with respect to $m$. We note that this can be resolved by using a race-track multi-pass configuration instead of a linear configuration as depicted in Fig. \ref{fig:setup}.

We calculate the SNR$_m/\sqrt{\left<\hat{n}\right>_\text{0}}$ at each pixel and its average value over a $120 \times \SI{30}{px}$ rectangle surrounding each cell center. 
The distributions of average SNR$_m/\sqrt{\left<\hat{n}\right>_\text{0}}$ for approximately 400 cells imaged at $m=1$, 2, and 4 are shown in Fig. \ref{fig:flow} along with Gaussian fits to the distributions. 
A linear fit to the mean SNR$_m/\sqrt{\left<\hat{n}\right>_\text{in}}$ as a function of $m$ yields a measured average transmission $\eta$ = 0.968 and an intercept resulting from phase contrast.
% consistent in magnitude with the contribution from phase contrast. \jrcomment{Do I need to be more specific about this "contribution"?} \yicomment{maybe resulting from phase contrast?}
The mean SNR$_m$ = 1.8, 2.9, and 4.0 at $m=1$, 2, and 4, respectively, which, as for the static case, shows a slight sub-linear enhancement due to loss in the optics. 
The increase in the distribution width as a function of $m$ results from the contrast enhancement and demonstrates the increased sensitivity of the multi-pass absorption measurement. 
For a transmission $\eta$ = $\eta_0$ $\pm$ $\delta \eta$, we find that the width of the distribution in $\text{SNR}/\sqrt{\left<\hat{n}\right>_\text{0}}$ should grow as $m \eta_0^{m-1}\delta\eta$ to first order in $\delta \eta$ and for high transmissivity, consistent with our measurement results. We further note that phase contrast and defocus through the flow channel can also contribute to the increased widths of the measured SNR distributions, however, these effects are minimized by selecting only in-focus cells for the analysis.

%\jrcomment{This seems a bit redundant with previous few paragraphs. Maybe drop or move the first two sentences into the previous paragraphs, and then say we could go even faster with higher illumination? It's a good transition at the end into the conclusion  anyway.} To limit motion-induced blur during flow to \SI{500}{\nano\meter}, comparable to the diffraction limit of our imaging system, for flow rates of up to \SI{50}{\milli\meter\per\second}, we operate at exposure times of \SI{10}{\micro\second}.
%In our imaging configuration, which uses a standard LED as the illumination source, this yields SNR$_1$ $\approx$ 1 for $m=1$ as shown in \autoref{fig:flow}.
Enhancing the SNR$_1$ $\approx$ 1 by a factor of two, as we do by going to $m=2$, at constant illumination would require a four-fold longer camera exposure, which is impossible without blurring the image or reducing the flow rate. 
Multi-pass provides the requisite enhancement in SNR to enable faster flow and thus higher cell throughput at constant exposure, and these gains are further enhanced for the $m=4$ configuration. 

\section{Conclusion}
%\jrcomment{Could include a sentence in here about extending to phase imaging with a thin sample and phase plate? Or no phase plate so phase and absorption can be distinguished by linear and quadratic scaling with m? Maybe also discuss absorption as a measurement of hemoglobin concentration and suggest this as a blood count diagnostic \cite{Kim2014,Kim2020,Mittman1997,Mohandas1986,GomezPastora2022}. Also could include label-free red blood cell/cancer applications here as they were cut from intro.}

We presented a multi-pass imaging technique capable of operating at two and four passes using continuous wave illumination and standard cameras. We applied our multi-pass scheme to imaging flow cytometry to image human red blood cells at increased sensitivity when limited by the illumination source intensity. We demonstrated SNR enhancement by up to a factor of four, under flow conditions comparable to conventional high-speed instruments. Alternatively, such an enhancement can be used to speed up acquisitions by a factor of up to 16 for similar sensitivity. Our multi-pass technique approaches close to the quantum limit for weak absorption measurements, while conventional (single-pass) absorption imaging is optimal for strong absorption, and both are achievable in our setup. Our technique presents a practical route for enhancing imaging sensitivity when the illumination is limited by the light source intensity, favoring multi-pass using classical light over proposed quantum light sources. Furthermore, it is easy to use and simple to implement, and thus could be widely deployed.

%This allows  increasing the signal-to-noise is agnostic to the source of noise.
%It is thus compatible with shot-noise limited detection arising from constrained illumination intensity, acquisition times, or the finite well depth of the detector.
%It can also offer an advantage when confronted with technical noise sources such as detector read noise.
%It can be used in conjunction with high well depth cameras to further leverage this advantage \cite{welldepth_ieee1,welldepth_ieee2,welldepth_patent,welldepth_adimec}.

%The method is compatible with other contrast enhancement methods, \emph{e.g.}, iSCAT or iSCAT-like techniques, by attenuating the reference beam outside of the cavity.
%A simple test was carried out with this apparatus by illuminating with a broadband laser and partially blocking the undiffracted (reference) beam in the Fourier plane of the extracavity imaging optics. We observed the expected contrast enhancement.
%The multi-pass method is also compatible with phase contrast (\emph{e.g.}, Zernike) and darkfield modalities, and could also be used in conjunction with techniques utilizing intracavity feedback elements \cite{lowphi}.
%For dose sensitive targets, multi-pass contrast gains can be exploited to reduce sample damage while maintaining image signal-to-noise \cite{theoryslog}.

%\begin{backmatter}
	\paragraph*{{Funding}} This work was done as part of the Quantum Electron Microscope collaboration funded by the Gordon and Betty Moore foundation.
	\paragraph*{Acknowledgments} We would like to thank Cheri M. Hampton, Lawrence F. Drummy, Adam J. Bowman, Dara P. Dowlatshahi, and Stewart Koppell for helpful discussions.
	\paragraph*{Disclosures} The authors declare no conflicts of interest.
	\paragraph*{Data availability} Data underlying the results presented in this paper are not publicly available at this time but may be obtained from the authors upon reasonable request.
%\end{backmatter}

\bigskip

%\bibliography{main}

\begin{thebibliography}{25}%
\makeatletter
\providecommand \@ifxundefined [1]{%
 \@ifx{#1\undefined}
}%
\providecommand \@ifnum [1]{%
 \ifnum #1\expandafter \@firstoftwo
 \else \expandafter \@secondoftwo
 \fi
}%
\providecommand \@ifx [1]{%
 \ifx #1\expandafter \@firstoftwo
 \else \expandafter \@secondoftwo
 \fi
}%
\providecommand \natexlab [1]{#1}%
\providecommand \enquote  [1]{``#1''}%
\providecommand \bibnamefont  [1]{#1}%
\providecommand \bibfnamefont [1]{#1}%
\providecommand \citenamefont [1]{#1}%
\providecommand \href@noop [0]{\@secondoftwo}%
\providecommand \href [0]{\begingroup \@sanitize@url \@href}%
\providecommand \@href[1]{\@@startlink{#1}\@@href}%
\providecommand \@@href[1]{\endgroup#1\@@endlink}%
\providecommand \@sanitize@url [0]{\catcode `\\12\catcode `\$12\catcode
  `\&12\catcode `\#12\catcode `\^12\catcode `\_12\catcode `\%12\relax}%
\providecommand \@@startlink[1]{}%
\providecommand \@@endlink[0]{}%
\providecommand \url  [0]{\begingroup\@sanitize@url \@url }%
\providecommand \@url [1]{\endgroup\@href {#1}{\urlprefix }}%
\providecommand \urlprefix  [0]{URL }%
\providecommand \Eprint [0]{\href }%
\providecommand \doibase [0]{https://doi.org/}%
\providecommand \selectlanguage [0]{\@gobble}%
\providecommand \bibinfo  [0]{\@secondoftwo}%
\providecommand \bibfield  [0]{\@secondoftwo}%
\providecommand \translation [1]{[#1]}%
\providecommand \BibitemOpen [0]{}%
\providecommand \bibitemStop [0]{}%
\providecommand \bibitemNoStop [0]{.\EOS\space}%
\providecommand \EOS [0]{\spacefactor3000\relax}%
\providecommand \BibitemShut  [1]{\csname bibitem#1\endcsname}%
\let\auto@bib@innerbib\@empty
%</preamble>
\bibitem [{\citenamefont {Hosseini}\ \emph {et~al.}(2016)\citenamefont
  {Hosseini}, \citenamefont {Zhou}, \citenamefont {Kim}, \citenamefont {Peres},
  \citenamefont {Diaspro}, \citenamefont {Kuang}, \citenamefont {Yaqoob},\ and\
  \citenamefont {So}}]{PhaseMicLimits2016}%
  \BibitemOpen
  \bibfield  {author} {\bibinfo {author} {\bibfnamefont {P.}~\bibnamefont
  {Hosseini}}, \bibinfo {author} {\bibfnamefont {R.}~\bibnamefont {Zhou}},
  \bibinfo {author} {\bibfnamefont {Y.-H.}\ \bibnamefont {Kim}}, \bibinfo
  {author} {\bibfnamefont {C.}~\bibnamefont {Peres}}, \bibinfo {author}
  {\bibfnamefont {A.}~\bibnamefont {Diaspro}}, \bibinfo {author} {\bibfnamefont
  {C.}~\bibnamefont {Kuang}}, \bibinfo {author} {\bibfnamefont
  {Z.}~\bibnamefont {Yaqoob}},\ and\ \bibinfo {author} {\bibfnamefont {P.~T.}\
  \bibnamefont {So}},\ }\bibfield  {title} {\bibinfo {title} {Pushing phase and
  amplitude sensitivity limits in interferometric microscopy},\ }\href@noop {}
  {\bibfield  {journal} {\bibinfo  {journal} {Optics Letters}\ }\textbf
  {\bibinfo {volume} {41}},\ \bibinfo {pages} {1656} (\bibinfo {year}
  {2016})}\BibitemShut {NoStop}%
\bibitem [{\citenamefont {Park}\ \emph {et~al.}(2018)\citenamefont {Park},
  \citenamefont {Depeursinge},\ and\ \citenamefont
  {Popescu}}]{PhasemicroscopyReview2018}%
  \BibitemOpen
  \bibfield  {author} {\bibinfo {author} {\bibfnamefont {Y.}~\bibnamefont
  {Park}}, \bibinfo {author} {\bibfnamefont {C.}~\bibnamefont {Depeursinge}},\
  and\ \bibinfo {author} {\bibfnamefont {G.}~\bibnamefont {Popescu}},\
  }\bibfield  {title} {\bibinfo {title} {Quantitative phase imaging in
  biomedicine},\ }\href@noop {} {\bibfield  {journal} {\bibinfo  {journal}
  {Nature Photonics}\ }\textbf {\bibinfo {volume} {12}},\ \bibinfo {pages}
  {578} (\bibinfo {year} {2018})}\BibitemShut {NoStop}%
\bibitem [{\citenamefont {Pirandola}\ \emph {et~al.}(2018)\citenamefont
  {Pirandola}, \citenamefont {Bardhan}, \citenamefont {Gehring}, \citenamefont
  {Weedbrook},\ and\ \citenamefont {Lloyd}}]{QsensReview2018}%
  \BibitemOpen
  \bibfield  {author} {\bibinfo {author} {\bibfnamefont {S.}~\bibnamefont
  {Pirandola}}, \bibinfo {author} {\bibfnamefont {B.~R.}\ \bibnamefont
  {Bardhan}}, \bibinfo {author} {\bibfnamefont {T.}~\bibnamefont {Gehring}},
  \bibinfo {author} {\bibfnamefont {C.}~\bibnamefont {Weedbrook}},\ and\
  \bibinfo {author} {\bibfnamefont {S.}~\bibnamefont {Lloyd}},\ }\bibfield
  {title} {\bibinfo {title} {Advances in photonic quantum sensing},\
  }\href@noop {} {\bibfield  {journal} {\bibinfo  {journal} {Nature Photonics}\
  }\textbf {\bibinfo {volume} {12}},\ \bibinfo {pages} {724} (\bibinfo {year}
  {2018})}\BibitemShut {NoStop}%
\bibitem [{\citenamefont {Moreau}\ \emph {et~al.}(2019)\citenamefont {Moreau},
  \citenamefont {Toninelli}, \citenamefont {Gregory},\ and\ \citenamefont
  {Padgett}}]{QimagReview2019}%
  \BibitemOpen
  \bibfield  {author} {\bibinfo {author} {\bibfnamefont {P.-A.}\ \bibnamefont
  {Moreau}}, \bibinfo {author} {\bibfnamefont {E.}~\bibnamefont {Toninelli}},
  \bibinfo {author} {\bibfnamefont {T.}~\bibnamefont {Gregory}},\ and\ \bibinfo
  {author} {\bibfnamefont {M.~J.}\ \bibnamefont {Padgett}},\ }\bibfield
  {title} {\bibinfo {title} {Imaging with quantum states of light},\
  }\href@noop {} {\bibfield  {journal} {\bibinfo  {journal} {Nature Reviews
  Physics}\ }\textbf {\bibinfo {volume} {1}},\ \bibinfo {pages} {367} (\bibinfo
  {year} {2019})}\BibitemShut {NoStop}%
\bibitem [{\citenamefont {Israel}\ \emph {et~al.}(2014)\citenamefont {Israel},
  \citenamefont {Rosen},\ and\ \citenamefont
  {Silberberg}}]{NOONmicroscopy2014}%
  \BibitemOpen
  \bibfield  {author} {\bibinfo {author} {\bibfnamefont {Y.}~\bibnamefont
  {Israel}}, \bibinfo {author} {\bibfnamefont {S.}~\bibnamefont {Rosen}},\ and\
  \bibinfo {author} {\bibfnamefont {Y.}~\bibnamefont {Silberberg}},\ }\bibfield
   {title} {\bibinfo {title} {Supersensitive polarization microscopy using noon
  states of light},\ }\href {https://doi.org/10.1103/PhysRevLett.112.103604}
  {\bibfield  {journal} {\bibinfo  {journal} {Phys. Rev. Lett.}\ }\textbf
  {\bibinfo {volume} {112}},\ \bibinfo {pages} {103604} (\bibinfo {year}
  {2014})}\BibitemShut {NoStop}%
\bibitem [{\citenamefont {Samantaray}\ \emph {et~al.}(2017)\citenamefont
  {Samantaray}, \citenamefont {Ruo-Berchera}, \citenamefont {Meda},\ and\
  \citenamefont {Genovese}}]{QlAbs2017}%
  \BibitemOpen
  \bibfield  {author} {\bibinfo {author} {\bibfnamefont {N.}~\bibnamefont
  {Samantaray}}, \bibinfo {author} {\bibfnamefont {I.}~\bibnamefont
  {Ruo-Berchera}}, \bibinfo {author} {\bibfnamefont {A.}~\bibnamefont {Meda}},\
  and\ \bibinfo {author} {\bibfnamefont {M.}~\bibnamefont {Genovese}},\
  }\bibfield  {title} {\bibinfo {title} {Realization of the first
  sub-shot-noise wide field microscope},\ }\href@noop {} {\bibfield  {journal}
  {\bibinfo  {journal} {Light: Science \& Applications}\ }\textbf {\bibinfo
  {volume} {6}},\ \bibinfo {pages} {e17005} (\bibinfo {year}
  {2017})}\BibitemShut {NoStop}%
\bibitem [{\citenamefont {Casacio}\ \emph {et~al.}(2021)\citenamefont
  {Casacio}, \citenamefont {Madsen}, \citenamefont {Terrasson}, \citenamefont
  {Waleed}, \citenamefont {Barnscheidt}, \citenamefont {Hage}, \citenamefont
  {Taylor},\ and\ \citenamefont {Bowen}}]{Warwick_NL_Nature2021}%
  \BibitemOpen
  \bibfield  {author} {\bibinfo {author} {\bibfnamefont {C.~A.}\ \bibnamefont
  {Casacio}}, \bibinfo {author} {\bibfnamefont {L.~S.}\ \bibnamefont {Madsen}},
  \bibinfo {author} {\bibfnamefont {A.}~\bibnamefont {Terrasson}}, \bibinfo
  {author} {\bibfnamefont {M.}~\bibnamefont {Waleed}}, \bibinfo {author}
  {\bibfnamefont {K.}~\bibnamefont {Barnscheidt}}, \bibinfo {author}
  {\bibfnamefont {B.}~\bibnamefont {Hage}}, \bibinfo {author} {\bibfnamefont
  {M.~A.}\ \bibnamefont {Taylor}},\ and\ \bibinfo {author} {\bibfnamefont
  {W.~P.}\ \bibnamefont {Bowen}},\ }\bibfield  {title} {\bibinfo {title}
  {Quantum-enhanced nonlinear microscopy},\ }\href@noop {} {\bibfield
  {journal} {\bibinfo  {journal} {Nature}\ }\textbf {\bibinfo {volume} {594}},\
  \bibinfo {pages} {201} (\bibinfo {year} {2021})}\BibitemShut {NoStop}%
\bibitem [{\citenamefont {Li}\ \emph {et~al.}(2022)\citenamefont {Li},
  \citenamefont {Li}, \citenamefont {Liu}, \citenamefont {Yakovlev},\ and\
  \citenamefont {Agarwal}}]{QSBS2022}%
  \BibitemOpen
  \bibfield  {author} {\bibinfo {author} {\bibfnamefont {T.}~\bibnamefont
  {Li}}, \bibinfo {author} {\bibfnamefont {F.}~\bibnamefont {Li}}, \bibinfo
  {author} {\bibfnamefont {X.}~\bibnamefont {Liu}}, \bibinfo {author}
  {\bibfnamefont {V.~V.}\ \bibnamefont {Yakovlev}},\ and\ \bibinfo {author}
  {\bibfnamefont {G.~S.}\ \bibnamefont {Agarwal}},\ }\bibfield  {title}
  {\bibinfo {title} {Quantum-enhanced stimulated brillouin scattering
  spectroscopy and imaging},\ }\href@noop {} {\bibfield  {journal} {\bibinfo
  {journal} {Optica}\ }\textbf {\bibinfo {volume} {9}},\ \bibinfo {pages} {959}
  (\bibinfo {year} {2022})}\BibitemShut {NoStop}%
\bibitem [{\citenamefont {Juffmann}\ \emph {et~al.}(2016)\citenamefont
  {Juffmann}, \citenamefont {Klopfer}, \citenamefont {Frankort}, \citenamefont
  {Haslinger},\ and\ \citenamefont {Kasevich}}]{mpm}%
  \BibitemOpen
  \bibfield  {author} {\bibinfo {author} {\bibfnamefont {T.}~\bibnamefont
  {Juffmann}}, \bibinfo {author} {\bibfnamefont {B.~B.}\ \bibnamefont
  {Klopfer}}, \bibinfo {author} {\bibfnamefont {T.~L.~I.}\ \bibnamefont
  {Frankort}}, \bibinfo {author} {\bibfnamefont {P.}~\bibnamefont
  {Haslinger}},\ and\ \bibinfo {author} {\bibfnamefont {M.~A.}\ \bibnamefont
  {Kasevich}},\ }\bibfield  {title} {{\bibinfo {title}
  {Multi-pass microscopy}},\ }\href {https://doi.org/10.1038/ncomms12858}
  {\bibfield  {journal} {\bibinfo  {journal} {Nature Communications}\ }\textbf
  {\bibinfo {volume} {7}},\ \bibinfo {pages} {1} (\bibinfo {year}
  {2016})}\BibitemShut {NoStop}%
\bibitem [{\citenamefont {Klopfer}\ \emph {et~al.}(2016)\citenamefont
  {Klopfer}, \citenamefont {Juffmann},\ and\ \citenamefont {Kasevich}}]{oam}%
  \BibitemOpen
  \bibfield  {author} {\bibinfo {author} {\bibfnamefont {B.~B.}\ \bibnamefont
  {Klopfer}}, \bibinfo {author} {\bibfnamefont {T.}~\bibnamefont {Juffmann}},\
  and\ \bibinfo {author} {\bibfnamefont {M.~A.}\ \bibnamefont {Kasevich}},\
  }\bibfield  {title} {{\bibinfo {title} {Iterative
  creation and sensing of twisted light}},\ }\href
  {https://doi.org/10.1364/OL.41.005744} {\bibfield  {journal} {\bibinfo
  {journal} {Optics Letters}\ }\textbf {\bibinfo {volume} {41}},\ \bibinfo
  {pages} {5744} (\bibinfo {year} {2016})}\BibitemShut {NoStop}%
\bibitem [{\citenamefont {Hur}\ \emph {et~al.}(2010)\citenamefont {Hur},
  \citenamefont {Tse},\ and\ \citenamefont {Carlo}}]{fast_flow00}%
  \BibitemOpen
  \bibfield  {author} {\bibinfo {author} {\bibfnamefont {S.~C.}\ \bibnamefont
  {Hur}}, \bibinfo {author} {\bibfnamefont {H.~T.~K.}\ \bibnamefont {Tse}},\
  and\ \bibinfo {author} {\bibfnamefont {D.~D.}\ \bibnamefont {Carlo}},\
  }\bibfield  {title} {{\bibinfo {title} {Sheathless
  inertial cell ordering for extreme throughput flow cytometry}},\ }\href
  {https://doi.org/10.1039/B919495A} {\bibfield  {journal} {\bibinfo  {journal}
  {Lab on a Chip}\ }\textbf {\bibinfo {volume} {10}},\ \bibinfo {pages} {274}
  (\bibinfo {year} {2010})}\BibitemShut {NoStop}%
\bibitem [{\citenamefont {Goda}\ \emph {et~al.}(2012)\citenamefont {Goda},
  \citenamefont {Ayazi}, \citenamefont {Gossett}, \citenamefont {Sadasivam},
  \citenamefont {Lonappan}, \citenamefont {Sollier}, \citenamefont {Fard},
  \citenamefont {Hur}, \citenamefont {Adam}, \citenamefont {Murray},
  \citenamefont {Wang}, \citenamefont {Brackbill}, \citenamefont {Carlo},\ and\
  \citenamefont {Jalali}}]{Goda2012}%
  \BibitemOpen
  \bibfield  {author} {\bibinfo {author} {\bibfnamefont {K.}~\bibnamefont
  {Goda}}, \bibinfo {author} {\bibfnamefont {A.}~\bibnamefont {Ayazi}},
  \bibinfo {author} {\bibfnamefont {D.~R.}\ \bibnamefont {Gossett}}, \bibinfo
  {author} {\bibfnamefont {J.}~\bibnamefont {Sadasivam}}, \bibinfo {author}
  {\bibfnamefont {C.~K.}\ \bibnamefont {Lonappan}}, \bibinfo {author}
  {\bibfnamefont {E.}~\bibnamefont {Sollier}}, \bibinfo {author} {\bibfnamefont
  {A.~M.}\ \bibnamefont {Fard}}, \bibinfo {author} {\bibfnamefont {S.~C.}\
  \bibnamefont {Hur}}, \bibinfo {author} {\bibfnamefont {J.}~\bibnamefont
  {Adam}}, \bibinfo {author} {\bibfnamefont {C.}~\bibnamefont {Murray}},
  \bibinfo {author} {\bibfnamefont {C.}~\bibnamefont {Wang}}, \bibinfo {author}
  {\bibfnamefont {N.}~\bibnamefont {Brackbill}}, \bibinfo {author}
  {\bibfnamefont {D.~D.}\ \bibnamefont {Carlo}},\ and\ \bibinfo {author}
  {\bibfnamefont {B.}~\bibnamefont {Jalali}},\ }\bibfield  {title} {\bibinfo
  {title} {High-throughput single-microparticle imaging flow analyzer},\ }\href
  {https://doi.org/10.1073/pnas.1204718109} {\bibfield  {journal} {\bibinfo
  {journal} {Proceedings of the National Academy of Sciences}\ }\textbf
  {\bibinfo {volume} {109}},\ \bibinfo {pages} {11630} (\bibinfo {year}
  {2012})},\ \BibitemShut {NoStop}%
\bibitem [{\citenamefont {Doan}\ \emph {et~al.}(2018)\citenamefont {Doan},
  \citenamefont {Vorobjev}, \citenamefont {Rees}, \citenamefont {Filby},
  \citenamefont {Wolkenhauer}, \citenamefont {Goldfeld}, \citenamefont
  {Lieberman}, \citenamefont {Barteneva}, \citenamefont {Carpenter},\ and\
  \citenamefont {Hennig}}]{Doan2018}%
  \BibitemOpen
  \bibfield  {author} {\bibinfo {author} {\bibfnamefont {M.}~\bibnamefont
  {Doan}}, \bibinfo {author} {\bibfnamefont {I.}~\bibnamefont {Vorobjev}},
  \bibinfo {author} {\bibfnamefont {P.}~\bibnamefont {Rees}}, \bibinfo {author}
  {\bibfnamefont {A.}~\bibnamefont {Filby}}, \bibinfo {author} {\bibfnamefont
  {O.}~\bibnamefont {Wolkenhauer}}, \bibinfo {author} {\bibfnamefont {A.~E.}\
  \bibnamefont {Goldfeld}}, \bibinfo {author} {\bibfnamefont {J.}~\bibnamefont
  {Lieberman}}, \bibinfo {author} {\bibfnamefont {N.}~\bibnamefont
  {Barteneva}}, \bibinfo {author} {\bibfnamefont {A.~E.}\ \bibnamefont
  {Carpenter}},\ and\ \bibinfo {author} {\bibfnamefont {H.}~\bibnamefont
  {Hennig}},\ }\bibfield  {title} {\bibinfo {title} {Diagnostic potential of
  imaging flow cytometry},\ }\href
  {https://doi.org/https://doi.org/10.1016/j.tibtech.2017.12.008} {\bibfield
  {journal} {\bibinfo  {journal} {Trends in Biotechnology}\ }\textbf {\bibinfo
  {volume} {36}},\ \bibinfo {pages} {649} (\bibinfo {year} {2018})}\BibitemShut
  {NoStop}%
\bibitem [{\citenamefont {Motsch}\ \emph {et~al.}(2010)\citenamefont {Motsch},
  \citenamefont {Zeppenfeld}, \citenamefont {Pinkse},\ and\ \citenamefont
  {Rempe}}]{motschCavityenhancedRayleighScattering2010}%
  \BibitemOpen
  \bibfield  {author} {\bibinfo {author} {\bibfnamefont {M.}~\bibnamefont
  {Motsch}}, \bibinfo {author} {\bibfnamefont {M.}~\bibnamefont {Zeppenfeld}},
  \bibinfo {author} {\bibfnamefont {P.~W.~H.}\ \bibnamefont {Pinkse}},\ and\
  \bibinfo {author} {\bibfnamefont {G.}~\bibnamefont {Rempe}},\ }\bibfield
  {title} {{\bibinfo {title} {Cavity-enhanced {{Rayleigh}}
  scattering}},\ }\href {https://doi.org/10.1088/1367-2630/12/6/063022}
  {\bibfield  {journal} {\bibinfo  {journal} {New Journal of Physics}\ }\textbf
  {\bibinfo {volume} {12}},\ \bibinfo {pages} {063022} (\bibinfo {year}
  {2010})}\BibitemShut {NoStop}%
\bibitem [{\citenamefont {Chang}\ \emph {et~al.}(2012)\citenamefont {Chang},
  \citenamefont {Jiang}, \citenamefont {Gorshkov},\ and\ \citenamefont
  {Kimble}}]{changCavityQEDAtomic2012}%
  \BibitemOpen
  \bibfield  {author} {\bibinfo {author} {\bibfnamefont {D.~E.}\ \bibnamefont
  {Chang}}, \bibinfo {author} {\bibfnamefont {L.}~\bibnamefont {Jiang}},
  \bibinfo {author} {\bibfnamefont {A.~V.}\ \bibnamefont {Gorshkov}},\ and\
  \bibinfo {author} {\bibfnamefont {H.~J.}\ \bibnamefont {Kimble}},\ }\bibfield
   {title} {{\bibinfo {title} {Cavity {{QED}} with atomic
  mirrors}},\ }\href {https://doi.org/10.1088/1367-2630/14/6/063003} {\bibfield
   {journal} {\bibinfo  {journal} {New Journal of Physics}\ }\textbf {\bibinfo
  {volume} {14}},\ \bibinfo {pages} {063003} (\bibinfo {year}
  {2012})}\BibitemShut {NoStop}%
\bibitem [{\citenamefont {Helstrom}(1976)}]{Helstrom}%
  \BibitemOpen
  \bibfield  {author} {\bibinfo {author} {\bibfnamefont {C.~W.}\ \bibnamefont
  {Helstrom}},\ }\href@noop {} {{\emph {\bibinfo
  {title} {Quantum Detection and Estimation Theory}}}}\ (\bibinfo  {publisher}
  {Academic Press},\ \bibinfo {year} {1976})\BibitemShut {NoStop}%
\bibitem [{\citenamefont {Holevo}(1982)}]{Holevo1982}%
  \BibitemOpen
  \bibfield  {author} {\bibinfo {author} {\bibfnamefont {A.~S.}\ \bibnamefont
  {Holevo}},\ }\href@noop {} {\emph {\bibinfo {title} {Probabilistic and
  Statistical Aspects of Quantum Theory}}}\ (\bibinfo  {publisher}
  {North-Holland Publishing Company},\ \bibinfo {year} {1982})\BibitemShut
  {NoStop}%
\bibitem [{\citenamefont {Braunstein}\ and\ \citenamefont
  {Caves}(1994)}]{Braunstein1994}%
  \BibitemOpen
  \bibfield  {author} {\bibinfo {author} {\bibfnamefont {S.~L.}\ \bibnamefont
  {Braunstein}}\ and\ \bibinfo {author} {\bibfnamefont {C.~M.}\ \bibnamefont
  {Caves}},\ }\bibfield  {title} {\bibinfo {title} {Statistical distance and
  the geometry of quantum states},\ }\href
  {https://doi.org/10.1103/PhysRevLett.72.3439} {\bibfield  {journal} {\bibinfo
   {journal} {Physical Review Letters}\ }\textbf {\bibinfo {volume} {72}},\
  \bibinfo {pages} {3439} (\bibinfo {year} {1994})}\BibitemShut {NoStop}%
\bibitem [{\citenamefont {Allen}\ \emph {et~al.}(2020)\citenamefont {Allen},
  \citenamefont {Sabines-Chesterking}, \citenamefont {McMillan}, \citenamefont
  {Joshi}, \citenamefont {Turner},\ and\ \citenamefont
  {Matthews}}]{JM_abs_PRR2020}%
  \BibitemOpen
  \bibfield  {author} {\bibinfo {author} {\bibfnamefont {E.~J.}\ \bibnamefont
  {Allen}}, \bibinfo {author} {\bibfnamefont {J.}~\bibnamefont
  {Sabines-Chesterking}}, \bibinfo {author} {\bibfnamefont {A.~R.}\
  \bibnamefont {McMillan}}, \bibinfo {author} {\bibfnamefont {S.~K.}\
  \bibnamefont {Joshi}}, \bibinfo {author} {\bibfnamefont {P.~S.}\ \bibnamefont
  {Turner}},\ and\ \bibinfo {author} {\bibfnamefont {J.~C.~F.}\ \bibnamefont
  {Matthews}},\ }\bibfield  {title} {\bibinfo {title} {Approaching the quantum
  limit of precision in absorbance estimation using classical resources},\
  }\href {https://doi.org/10.1103/PhysRevResearch.2.033243} {\bibfield
  {journal} {\bibinfo  {journal} {Phys. Rev. Research}\ }\textbf {\bibinfo
  {volume} {2}},\ \bibinfo {pages} {033243} (\bibinfo {year}
  {2020})}\BibitemShut {NoStop}%
\bibitem [{\citenamefont {Birchall}\ \emph {et~al.}(2020)\citenamefont
  {Birchall}, \citenamefont {Allen}, \citenamefont {Stace}, \citenamefont
  {O'Brien}, \citenamefont {Matthews},\ and\ \citenamefont
  {Cable}}]{JM_PhaseAbs_PRL2020}%
  \BibitemOpen
  \bibfield  {author} {\bibinfo {author} {\bibfnamefont {P.~M.}\ \bibnamefont
  {Birchall}}, \bibinfo {author} {\bibfnamefont {E.~J.}\ \bibnamefont {Allen}},
  \bibinfo {author} {\bibfnamefont {T.~M.}\ \bibnamefont {Stace}}, \bibinfo
  {author} {\bibfnamefont {J.~L.}\ \bibnamefont {O'Brien}}, \bibinfo {author}
  {\bibfnamefont {J.~C.~F.}\ \bibnamefont {Matthews}},\ and\ \bibinfo {author}
  {\bibfnamefont {H.}~\bibnamefont {Cable}},\ }\bibfield  {title} {\bibinfo
  {title} {Quantum optical metrology of correlated phase and loss},\ }\href
  {https://doi.org/10.1103/PhysRevLett.124.140501} {\bibfield  {journal}
  {\bibinfo  {journal} {Phys. Rev. Lett.}\ }\textbf {\bibinfo {volume} {124}},\
  \bibinfo {pages} {140501} (\bibinfo {year} {2020})}\BibitemShut {NoStop}%
\bibitem [{\citenamefont {Slussarenko}\ \emph {et~al.}(2017)\citenamefont
  {Slussarenko}, \citenamefont {Weston}, \citenamefont {Chrzanowski},
  \citenamefont {Shalm}, \citenamefont {Verma}, \citenamefont {Nam},\ and\
  \citenamefont {Pryde}}]{NOONoverSNL_NPh2017}%
  \BibitemOpen
  \bibfield  {author} {\bibinfo {author} {\bibfnamefont {S.}~\bibnamefont
  {Slussarenko}}, \bibinfo {author} {\bibfnamefont {M.~M.}\ \bibnamefont
  {Weston}}, \bibinfo {author} {\bibfnamefont {H.~M.}\ \bibnamefont
  {Chrzanowski}}, \bibinfo {author} {\bibfnamefont {L.~K.}\ \bibnamefont
  {Shalm}}, \bibinfo {author} {\bibfnamefont {V.~B.}\ \bibnamefont {Verma}},
  \bibinfo {author} {\bibfnamefont {S.~W.}\ \bibnamefont {Nam}},\ and\ \bibinfo
  {author} {\bibfnamefont {G.~J.}\ \bibnamefont {Pryde}},\ }\bibfield  {title}
  {\bibinfo {title} {Unconditional violation of the shot-noise limit in
  photonic quantum metrology},\ }\href@noop {} {\bibfield  {journal} {\bibinfo
  {journal} {Nature Photonics}\ }\textbf {\bibinfo {volume} {11}},\ \bibinfo
  {pages} {700} (\bibinfo {year} {2017})}\BibitemShut {NoStop}%
\bibitem [{\citenamefont {Belsley}\ \emph {et~al.}(2022)\citenamefont
  {Belsley}, \citenamefont {Allen}, \citenamefont {Datta},\ and\ \citenamefont
  {Matthews}}]{JM_cavity_PRL2022}%
  \BibitemOpen
  \bibfield  {author} {\bibinfo {author} {\bibfnamefont {A.}~\bibnamefont
  {Belsley}}, \bibinfo {author} {\bibfnamefont {E.~J.}\ \bibnamefont {Allen}},
  \bibinfo {author} {\bibfnamefont {A.}~\bibnamefont {Datta}},\ and\ \bibinfo
  {author} {\bibfnamefont {J.~C.}\ \bibnamefont {Matthews}},\ }\bibfield
  {title} {\bibinfo {title} {Advantage of coherent states in ring resonators
  over any quantum probe single-pass absorption estimation strategy},\
  }\href@noop {} {\bibfield  {journal} {\bibinfo  {journal} {Physical Review
  Letters}\ }\textbf {\bibinfo {volume} {128}},\ \bibinfo {pages} {230501}
  (\bibinfo {year} {2022})}\BibitemShut {NoStop}%
\bibitem [{\citenamefont {Birchall}\ \emph {et~al.}(2017)\citenamefont
  {Birchall}, \citenamefont {O'Brien}, \citenamefont {Matthews},\ and\
  \citenamefont {Cable}}]{JM_LossyPhase_PRA2017}%
  \BibitemOpen
  \bibfield  {author} {\bibinfo {author} {\bibfnamefont {P.~M.}\ \bibnamefont
  {Birchall}}, \bibinfo {author} {\bibfnamefont {J.~L.}\ \bibnamefont
  {O'Brien}}, \bibinfo {author} {\bibfnamefont {J.~C.}\ \bibnamefont
  {Matthews}},\ and\ \bibinfo {author} {\bibfnamefont {H.}~\bibnamefont
  {Cable}},\ }\bibfield  {title} {\bibinfo {title} {Quantum-classical boundary
  for precision optical phase estimation},\ }\href@noop {} {\bibfield
  {journal} {\bibinfo  {journal} {Physical Review A}\ }\textbf {\bibinfo
  {volume} {96}},\ \bibinfo {pages} {062109} (\bibinfo {year}
  {2017})}\BibitemShut {NoStop}%
\bibitem [{\citenamefont {Koppell}\ and\ \citenamefont
  {Kasevich}(2022)}]{koppell2022Arxivl}%
  \BibitemOpen
  \bibfield  {author} {\bibinfo {author} {\bibfnamefont {S.~A.}\ \bibnamefont
  {Koppell}}\ and\ \bibinfo {author} {\bibfnamefont {M.~A.}\ \bibnamefont
  {Kasevich}},\ }\bibfield  {title} {\bibinfo {title} {Optimal dose-limited
  phase estimation without entanglement},\ }\href@noop {} {\bibfield  {journal}
  {\bibinfo  {journal} {arXiv preprint arXiv:2203.10137}\ } (\bibinfo {year}
  {2022})}\BibitemShut {NoStop}%
\bibitem [{Ima()}]{ImageStream}%
  \BibitemOpen
  \href@noop {} {}\bibinfo {howpublished} {See for example, Amnis
  ImageStream$^X$ Mk II,
  https://www.luminexcorp.com/imagestreamx-mk-ii}\BibitemShut {NoStop}%
\end{thebibliography}

\begin{thebibliography}{3}%
\makeatletter
\providecommand \@ifxundefined [1]{%
 \@ifx{#1\undefined}
}%
\providecommand \@ifnum [1]{%
 \ifnum #1\expandafter \@firstoftwo
 \else \expandafter \@secondoftwo
 \fi
}%
\providecommand \@ifx [1]{%
 \ifx #1\expandafter \@firstoftwo
 \else \expandafter \@secondoftwo
 \fi
}%
\providecommand \natexlab [1]{#1}%
\providecommand \enquote  [1]{``#1''}%
\providecommand \bibnamefont  [1]{#1}%
\providecommand \bibfnamefont [1]{#1}%
\providecommand \citenamefont [1]{#1}%
\providecommand \href@noop [0]{\@secondoftwo}%
\providecommand \href [0]{\begingroup \@sanitize@url \@href}%
\providecommand \@href[1]{\@@startlink{#1}\@@href}%
\providecommand \@@href[1]{\endgroup#1\@@endlink}%
\providecommand \@sanitize@url [0]{\catcode `\\12\catcode `\$12\catcode
  `\&12\catcode `\#12\catcode `\^12\catcode `\_12\catcode `\%12\relax}%
\providecommand \@@startlink[1]{}%
\providecommand \@@endlink[0]{}%
\providecommand \url  [0]{\begingroup\@sanitize@url \@url }%
\providecommand \@url [1]{\endgroup\@href {#1}{\urlprefix }}%
\providecommand \urlprefix  [0]{URL }%
\providecommand \Eprint [0]{\href }%
\providecommand \doibase [0]{https://doi.org/}%
\providecommand \selectlanguage [0]{\@gobble}%
\providecommand \bibinfo  [0]{\@secondoftwo}%
\providecommand \bibfield  [0]{\@secondoftwo}%
\providecommand \translation [1]{[#1]}%
\providecommand \BibitemOpen [0]{}%
\providecommand \bibitemStop [0]{}%
\providecommand \bibitemNoStop [0]{.\EOS\space}%
\providecommand \EOS [0]{\spacefactor3000\relax}%
\providecommand \BibitemShut  [1]{\csname bibitem#1\endcsname}%
\let\auto@bib@innerbib\@empty
%</preamble>
\bibitem [{\citenamefont {Birchall}\ \emph {et~al.}(2020)\citenamefont
  {Birchall}, \citenamefont {Allen}, \citenamefont {Stace}, \citenamefont
  {O'Brien}, \citenamefont {Matthews},\ and\ \citenamefont
  {Cable}}]{JM_PhaseAbs_PRL2020}%
  \BibitemOpen
  \bibfield  {author} {\bibinfo {author} {\bibfnamefont {P.~M.}\ \bibnamefont
  {Birchall}}, \bibinfo {author} {\bibfnamefont {E.~J.}\ \bibnamefont {Allen}},
  \bibinfo {author} {\bibfnamefont {T.~M.}\ \bibnamefont {Stace}}, \bibinfo
  {author} {\bibfnamefont {J.~L.}\ \bibnamefont {O'Brien}}, \bibinfo {author}
  {\bibfnamefont {J.~C.~F.}\ \bibnamefont {Matthews}},\ and\ \bibinfo {author}
  {\bibfnamefont {H.}~\bibnamefont {Cable}},\ }\bibfield  {title} {\bibinfo
  {title} {Quantum optical metrology of correlated phase and loss},\ }\href
  {https://doi.org/10.1103/PhysRevLett.124.140501} {\bibfield  {journal}
  {\bibinfo  {journal} {Phys. Rev. Lett.}\ }\textbf {\bibinfo {volume} {124}},\
  \bibinfo {pages} {140501} (\bibinfo {year} {2020})}\BibitemShut {NoStop}%
\bibitem [{\citenamefont {Birchall}\ \emph {et~al.}(2017)\citenamefont
  {Birchall}, \citenamefont {O'Brien}, \citenamefont {Matthews},\ and\
  \citenamefont {Cable}}]{JM_LossyPhase_PRA17}%
  \BibitemOpen
  \bibfield  {author} {\bibinfo {author} {\bibfnamefont {P.~M.}\ \bibnamefont
  {Birchall}}, \bibinfo {author} {\bibfnamefont {J.~L.}\ \bibnamefont
  {O'Brien}}, \bibinfo {author} {\bibfnamefont {J.~C.~F.}\ \bibnamefont
  {Matthews}},\ and\ \bibinfo {author} {\bibfnamefont {H.}~\bibnamefont
  {Cable}},\ }\bibfield  {title} {\bibinfo {title} {Quantum-classical boundary
  for precision optical phase estimation},\ }\href
  {https://doi.org/10.1103/PhysRevA.96.062109} {\bibfield  {journal} {\bibinfo
  {journal} {Phys. Rev. A}\ }\textbf {\bibinfo {volume} {96}},\ \bibinfo
  {pages} {062109} (\bibinfo {year} {2017})}\BibitemShut {NoStop}%
\bibitem [{\citenamefont {Corless}\ \emph {et~al.}(1996)\citenamefont
  {Corless}, \citenamefont {Gonnet}, \citenamefont {Hare}, \citenamefont
  {Jeffrey},\ and\ \citenamefont {Knuth}}]{Lambert}%
  \BibitemOpen
  \bibfield  {author} {\bibinfo {author} {\bibfnamefont {R.~M.}\ \bibnamefont
  {Corless}}, \bibinfo {author} {\bibfnamefont {G.~H.}\ \bibnamefont {Gonnet}},
  \bibinfo {author} {\bibfnamefont {D.~E.}\ \bibnamefont {Hare}}, \bibinfo
  {author} {\bibfnamefont {D.~J.}\ \bibnamefont {Jeffrey}},\ and\ \bibinfo
  {author} {\bibfnamefont {D.~E.}\ \bibnamefont {Knuth}},\ }\bibfield  {title}
  {\bibinfo {title} {On the {L}ambert {W} function},\ }\href@noop {} {\bibfield
   {journal} {\bibinfo  {journal} {Advances in Computational Mathematics}\
  }\textbf {\bibinfo {volume} {5}},\ \bibinfo {pages} {329} (\bibinfo {year}
  {1996})}\BibitemShut {NoStop}%
\end{thebibliography}
%\bibliographyfullrefs{main}
%

%%%%%%%%%% Merge with supplemental materials %%%%%%%%%%
%\pagebreak
\widetext
%%%%%%%%%% Merge with supplemental materials %%%%%%%%%%
%%%%%%%%%% Prefix a "S" to all equations, figures, tables and reset the counter %%%%%%%%%%
\setcounter{equation}{0}
\setcounter{figure}{0}
\setcounter{table}{0}
\setcounter{page}{1}
\setcounter{section}{0}
\makeatletter
\renewcommand{\theequation}{S\arabic{equation}}
\renewcommand{\thefigure}{S\arabic{figure}}
\renewcommand{\bibnumfmt}[1]{[S#1]}
\renewcommand{\citenumfont}[1]{S#1}
%%%%%%%%%% Prefix a "S" to all equations, figures, tables and reset the counter %%%%%%%%%%

\begin{center}
\textbf{\large Continuous wave multipass imaging: supplemental document}
\end{center}

%%% BEGIN SUPP FOR INCLUDING IN MAIN ARIX .tex
\section{Bright-field Multi-pass Absorption and Phase Contrast}
The number of detected photons in bright-field multi-pass imaging of an object after $m$ interactions is 
%\begin{equation}
    \begin{align}
        \left< \hat{n} \right> & = \left<\hat{\tilde{n}}\right>_\text{in} \eta^m \eta_\text{rt}^m \eta_D |e^{\imath\theta}+ s e^{\imath m\phi}|^2 \nonumber \\
        & = \left<\hat{\tilde{n}}\right>_\text{in} \eta^m \eta_\text{rt}^m \eta_D (1+ 2 s \cos(m\phi-\theta) + s^{2}) ,
    \label{eq:n_D}
    \end{align}
%\end{equation}
where $\eta$ is the transmission of the specimen, $\phi$ is the specimen phase shift, $\eta_\text{rt}$ is the round-trip optical loss, $\eta_D$ is the detection loss, $\theta$ is an auxiliary phase applied to the reference field (e.g. $\theta=\pi/2$ for a Zernike phase plate), $s$ is the object scattered amplitude after $m$ interactions, and $\left<\hat{\tilde{n}}\right>_\text{in} \equiv \left<\hat{n}\right>_\text{in}/(|e^{\imath \theta}+s|^2)$, where $\left<\hat{n}\right>_\text{in}$ is the average number of photons at the apparatus input. The signal-to-noise ratio (SNR) is hence 
\begin{align}
    \text{SNR}_m %& \approx \frac{|\left< \hat{n} \right>-\eta_\text{rt}^m\left< \hat{n} \right>_\text{in}|}{\sqrt{\left< \hat{n} \right>+\eta_\text{rt}^m\left< \hat{n} \right>_\text{in}}} \\
    & \approx \frac{\left|\left< \hat{n} \right>-\left< \hat{n} \right>_{\eta=1}\right|}{\sqrt{\left< \hat{n} \right>+\left< \hat{n} \right>_{\eta=1}}} \\
    & = \frac{\sqrt{\eta_\text{rt}^m\eta_D\left< \hat{n} \right>_\text{in}}}{|e^{\imath \theta}+s|}\frac{2s(\cos(\theta)-\cos(m\phi-\theta))+(1+ 2 s \cos(m\phi-\theta) + s^{2})(1-\eta^m)}{\sqrt{(1+ s^{2})(1+ \eta^m) + 2s(\cos(\theta)+ \eta^m \cos(m\phi-\theta))}} \nonumber.
\end{align}
In the limit of a weak phase, weakly absorbing and scattering object, the SNR can be simplified. We drop terms of $O(s^2)$ and expand to first non-zero order in $\phi$ and the absorption, written in terms of the absorption per length $\alpha$ and the sample thickness $z$, such that $\eta = e^{-\alpha z}$. The result is

\begin{align}
    & \text{SNR}_m\sim \sqrt{\frac{\left<\hat{n}\right>_0}{2}} [s\phi^2m^2 +\alpha z m ] \,\,\,\, (for \,\,\theta=0),\\
    & \text{SNR}_m\sim \sqrt{\frac{\left<\hat{n}\right>_0}{2}} [2s\phi m +\alpha z m ] \,\,\,\,\, (for \,\,\theta=\pi/2),
\end{align}
where the average squared-root photon flux without the object $\left<\hat{n}\right>_0$ $\equiv$ $\eta_\text{rt}^m\eta_D\left<\hat{n}\right>_\text{in}$. 
%Note that in our setup, for $m=1$, additional losses occur due to defocusing objective O1, yielding an effective $\eta_\text{rt}$ for $m=1$. 
%\jrcomment{Note that in our setup, the optical path changes between $m=1$, 2, and 4, resulting in an effective $\eta_\text{rt}$ for each number of passes that does not build up as $\eta_\text{rt}^m$.}
Note that in our implementation of the multi-pass setup, the optical path changes between $m=1$, 2, and 4, resulting in an effective $\eta_\text{rt}$ for each number of passes that does not build up as $\eta_\text{rt}^m$. This is particularly due to additional losses in defocusing objective O1 and optical elements, yielding a larger effective $\eta_\text{rt}$ for $m=1$, as well as the additional optical path for $m=4$.

%The quantum Fisher information (QFI) per input illumination photon number for multi-passed classical light is given by
%\begin{equation}
 %   \tilde{\mathcal{F}}_m = m^2(T\eta)^{m-2},
%\end{equation}
%and the maximal multi-pass QFI becomes $\tilde{\mathcal{F}}_{m_{\text{opt}}} = 4/(e\cdot T \eta\ln{(T \eta)})^2$ for $m_{\text{opt}} = -2/\ln{(T \eta)}$.
 
\section{Quantum limit for absorption estimation}
 In this section we give more details on the calculation of the QFI limit to absorption estimation (Equation 6 from the main text), denoting the maximum information available on $\eta$ for any quantum probe and measurement which includes multiple interactions of the probe with the sample. The QFI for absorption estimation with quantum probes and $m$ passes is given by \cite{JM_PhaseAbs_PRL2020}
\begin{align}
    \QFI_{Q,m} = \langle \hat{n} \rangle_{\text{in}} m^2\eta^{m-2}/(1-\eta^m), \label{eq:QFI_QM}
\end{align}
where $\langle \hat{n} \rangle_{\text{in}}$ is the mean total number of photons in the input probe state $\rho_\text{in}$. An expression similar to Eq. \ref{eq:QFI_QM} was optimized in Ref. \cite{JM_LossyPhase_PRA17} when $\tilde{m}_{\text{opt}}= -(2+\mathcal{W})/\ln{\eta}\approx-1.59/\ln{\eta}$, where $\mathcal{W}\equiv W_0(-2/e^2)\approx-0.406$ is the main branch of the Lambert $W$ function at $-2/e^2$ \cite{Lambert}. Inserting $\tilde{m}_{\text{opt}}$ into Eq. \ref{eq:QFI_QM} we find the quantum limit to absorption estimation limited by number of input probe photons $\QFIn_Q=\QFI_{Q,\tilde{m}_{\text{opt}}}/\langle \hat{n} \rangle_{\text{in} }$ (Eq. 6, main text),
\begin{align}
    \QFIn_Q =-\frac{\mathcal{W}(2+\mathcal{W})}{(\eta\ln{\eta})^2}\approx
    \frac{0.65}{(\eta\ln{\eta})^2}.    
    \label{eq:QFI_Q}
\end{align}
 
\section{Saturating the Cramer-Rao Bound for Multi-pass loss sensitivity}
The uncertainty in $\eta$ can be estimated from equation \ref{eq:n_D} using simple error propagation,
\begin{align}
    & \Delta \eta = \left(\frac{\partial \langle \hat{n} \rangle}{\partial \eta} \right)^{-1} \Delta \langle \hat{n} \rangle_ 
    = \left(\langle \hat{n} \rangle_{in} m \eta^{m-1}\eta_{rt}^m\eta_D \right)^{-1} \Delta \langle \hat{n} \rangle,
\end{align}
where, for absorption contrast, we take $\phi = \theta = 0$. Taking $\eta_D=1$ and $\eta_{rt} =1$, for a shot-noise limited measurement, $\Delta \langle \hat{n} \rangle$ = $\sqrt{\langle \hat{n} \rangle_{in} \eta^m}$, and the inverse of the variance in $\eta$ saturates the Cramer-Rao bound, 
\begin{equation}
     \frac{1}{\text{Var}(\eta)} = \langle \hat{n} \rangle_{in} m^2 \eta^{m-2} = \mathcal{F}_m . 
\end{equation}

\end{document}